\documentclass[conf]{new-aiaa}
\usepackage[utf8]{inputenc}

\usepackage{graphicx}
\usepackage{amsmath}
\DeclareMathOperator*{\argmin}{arg\,min}  
\usepackage[version=4]{mhchem}
\usepackage{siunitx}
\usepackage{longtable,tabularx}
\usepackage{caption}
\usepackage{subcaption}
\usepackage[ruled]{algorithm2e}
\usepackage{comment}

\title{Active Learning Enhanced Surrogate Modeling of Jet Engines in JuliaSim}

\author{Anas Abdelrehim*, Dhairya Gandhi\footnote{These authors contributed equally.}, Sharan Yalburgi, Ashutosh Bharambe, Ranjan Anantharaman, Chris Rackauckas}
\affil{JuliaHub Inc., Cambridge, MA, 02138}

\begin{document}

\maketitle

\begin{abstract}
Surrogate models are effective tools for accelerated design of complex systems. The result of a design optimization procedure using surrogate models can be used to initialize an optimization routine using the full order system. High accuracy of the surrogate model can be advantageous for fast convergence. In this work, we present an active learning approach to produce a very high accuracy surrogate model of a turbofan jet engine, that demonstrates 0.1\% relative error for all quantities of interest. We contrast this with a surrogate model produced using a more traditional brute-force data generation approach. 

\end{abstract}

\section{Nomenclature \& Units}

{\renewcommand\arraystretch{1.0}
\noindent\begin{longtable*}{@{}l @{\quad=\quad} l@{}}
$\text{Amb.alt\_in}$  & pressure altitude (in feet) \\
$\text{Amb.dTs\_in}$ & deviation from standard ambient temperature (R)\\
$\text{Amb.MN\_in}$ & free stream Mach Number (unitless) \\
$\text{WfmSet.NL}$ & desired shaft speed (RPM) \\
ShH\_N & mechanical speed of the shaft (RPM) \\
F049\_Tt & Turbine Temperature (R) \\
PerfInst\_Fn & Net Thrust (lbf) \\
PerfInst\_SFC & Specific Fuel Consumption (unitless)\\
PerfInst\_WfuelPPH & Net Fuel Flow (lbm/hr)\\

\end{longtable*}}

\section{Introduction}
\lettrine{S}{urrogate} modeling is a popular method in the aerospace community for accelerated aerodynamic design \cite{yondo2018review, tao2019application}. Detailed computer simulations of real-world systems can be computationally expensive, and analyses such as uncertainty quantification and design optimization may require thousands of run to yield a solution \cite{schaefer2020approaches, beran2017uncertainty}. In the aerospace community, various surrogate modeling methods such as response surfaces \cite{jansson2003using}, Kriging \cite{laurenceau2008building}, Radial Basis Functions \cite{jakobsson2007mesh} and Deep Neural Networks \cite{li2019deep} have been explored. 

\begin{figure}[hbt!]
\centering
\includegraphics[width=.9\textwidth]{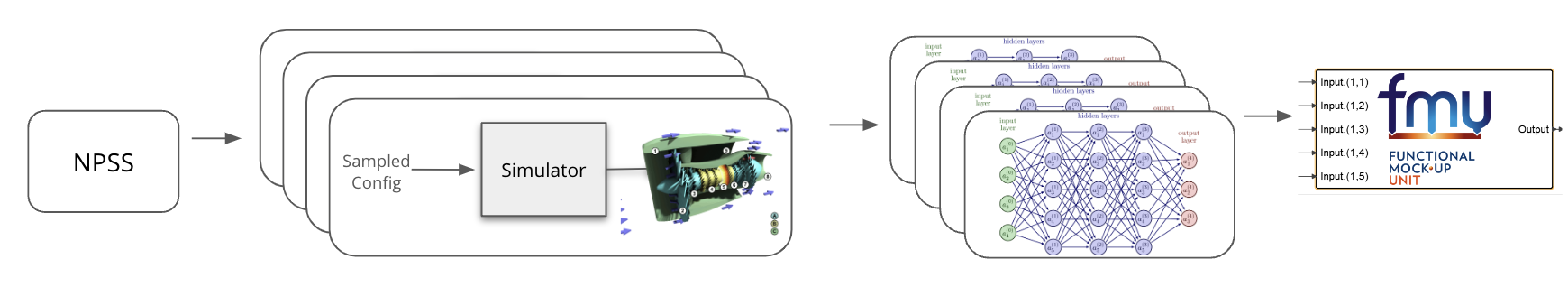}
\caption{Surrogate model generation pipeline in JuliaSim}
\end{figure}

An effective approach for system design optimization is to use a surrogate model to initialize the design optimization routine \cite{keane2020robust}. In this approach, a surrogate model is trained to replicate a dynamic mapping between the design variables and the dynamics of the system, with an input space representative of the design search space \cite{mack2007surrogate, williams2019surrogate}. A design optimization is then conducted using the surrogate model, the result of which is to initialize the design optimization of the reference system. The rationale is that the original design problem would be initialized closed to its minima, leading to fast convergence \cite{wortmann2015advantages, mack2007surrogate, han2012surrogate}. To this end, a surrogate model with high accuracy can ensure that the surrogate guided design optimization converges somewhere close to the desired minima \cite{alizadeh2020managing, li2019surrogate}.

In this work, we demonstrate an active learning methodology for generation of a high accuracy surrogate model of a turbofan jet engine, and contrast it with a brute-force approach based on quasi-monte carlo sampling. The reference physics-based model of a turbo-fan jet engine was built using NPSS \cite{lytle2000numerical}, maintained by the Southwest Research Institute. It consists of standard thermodynamic process models, which are used to construct a functional aero-thermal performance model of a two-spool turbofan system. The flight envelope over which it can successfully operate is described as function of the International Standard Atmosphere (ISA), and is given by the Tables \ref{tab:fe1} and \ref{tab:fe2}. For this study, we utilize the surrogate modeling method known as the digital echo \cite{rackauckasscientific}, which is part of JuliaSim \cite{rackauckas2022composing}, a modeling and simulation framework in the Julia programming language \cite{bezanson2017julia}. JuliaSim's surrogate modeling algorithms have been demonstrated on district heating systems \cite{anantharaman2020accelerating}, systems pharmacology \cite{anantharaman2022stably}, room air conditioner models in buildings \cite{rackauckas2022composing}, digital circuits \cite{anantharaman2021composable} and power systems in the literature \cite{roberts2022continuous}. 

The rest of this paper is organized as follows: in the following section, we shall introduce the methods, namely the surrogate modeling and active learning methods, followed by the results before and after the active learning was applied. We also present some statistics on the error and show that almost all of the error is concentrated on the boundary of the domain. 

\begin{table}
\caption{\label{tab:table1} Flight envelope of Turbofan model: pressure altitude, ambient temperature and shaft speed}
\centering
\begin{tabular}{lccc}
\hline
Quantity & Lower bound & Upper bound & Units \\\hline
Amb.alt\_in & -2000 & 10000 & feet \\
Amb.dTs\_in & -35 & 35 & deg F delta ISA (dTs) \\
WfmSet.NL & 2000 & 5000  & RPM\\
\hline
\end{tabular}
\label{tab:fe1}
\end{table}

\begin{table}
\caption{\label{tab:table1} Flight envelope of Turbofan model: Mach Number}
\centering
\begin{tabular}{lcc}
\hline
Altitude (ft) & Lower bound & Upper bound \\\hline
-2000 & 0 & 0.5 \\
10000 & 0 & 0.58 \\
20000 & 0.4 & 0.68\\
30000 & 0.4 & 0.9\\
40000 & 0.4 & 0.9\\
\hline
\end{tabular}
\label{tab:fe2}
\end{table}

\section{Methods}

\subsection{Surrogate Modeling Method - the Digital Echo}
The digital echo surrogate generation pipeline is a data-driven framework which constructs a new steady state system $S$ from a full-order steady state system $M$. The surrogate model system $S$ is represented by: 

\begin{align*}
x^\prime &= 0\\
f(p) &= x
\end{align*}

where $p$ represents the model parameter (which is consistent with the full-order system).

The Digital Echo is trained by first sampling the input space of the physics model before running the physics model to generate training data at the sampled points. The training procedure is represented by the following sequence:

\begin{enumerate}
    \item Establish error criteria the surrogate generation pipeline must converge to.
    \item Sample input space using a Quasi-Monte Carlo scheme to generate a set of flight parameters $\{p_1, \dots, p_{n_p} \}$ from the flight envelope $P$. 
    \item Simulate the physics-based model at each parameter, to generate a set of results $\{r_1, \dots, r_{n_r}\}$. 
    \item Split dataset into two subsets by a fixed split ratio. These sets, along with the corresponding parameters represent the training and the validation set respectively. 
    \item Generate a set of hyperparameters $\{h_1, \dots, h_{n_h}\} \in H$ where $H$ represents the hyperparameter space for the Digital Echo.
    \item For every $h_i \in H$, use the Digital Echo to generate surrogate model $s_i$
    \item Evaluate the model(s) on the test set, take the best performing hyperparameters and perform active learning to redo steps 2-6 until convergence to desired error criteria.
\end{enumerate}

The above pipeline is constructed to support the generation of high (or low) accuracy surrogate models, depending on the compute budget and application. 

\SetKwComment{Comment}{/* }{ */}
\begin{algorithm}
\caption{Adaptive Downsampling Routine}\label{alg:al1}
\KwData{Input parameter space $P: (P_{LB}, P_{UB})$, output variable set $Y$, Simulator $\Psi$}
\KwResult{Surrogate Model $\Phi: P \mapsto Y$}
$P_N \gets QMC(P_{LB}, P_{UB}; N)$\ \Comment*[r]{Densely sample input space to extract N samples}
$T_N \gets \Psi(P_N)$\ \Comment*[r]{Simulate at each sample to generate training set}
$Y_{UB} \gets \max_{T_N} Y$\ \Comment*[r]{Compute upper and lower bounds for the outputs}
$Y_{LB} \gets \min_{T_N} Y$\;
$Y_N \gets QMC(Y_{LB}, Y_{UB}; N)$\ \Comment*[r]{Densely sample the output}
$(T_n, P_n) \gets \texttt{NearestNeighbor}(Y_N, T_N, P_N) $\ \Comment*[r]{Downsample to n points}
$\Phi \gets DE(T_n, P_n)$;\ \Comment*[r]{Use improved training set to generate surrogate model}
\end{algorithm}

\begin{algorithm}
\caption{Nearest Neighbor Computation and Downsampling (\texttt{NearestNeighbor})}\label{alg:al2}
\KwData{Sampled outputs $Y_N$, Training Set $T_N$, Parameter Set $P_N$}
\KwResult{Downsampled training set $T_n$, Downsampled Input Samples $P_n$}
$i \gets 1$\;
$T_n \gets \phi$\ \Comment*[r]{Initialize to empty set}
$P_n \gets \phi$\ \Comment*[r]{Initialize to empty set}
\For{i < N}{
    $t \gets \argmin_{T_j \in T_N} dist(y_i, T_j)$\ \Comment*[r]{Distance metric used is the manhattan distance}
    $T_n \gets T_n \cup \{t\}$\;
    $P_n \gets P_n \cup {p}$\ \Comment*[r]{where $\Psi(p) = t$}
}
$T_n \gets \texttt{unique}(T_n)$\;
\end{algorithm}

\subsection{Active Learning}

\begin{figure}
    \centering
    \includegraphics[width=0.7\linewidth]{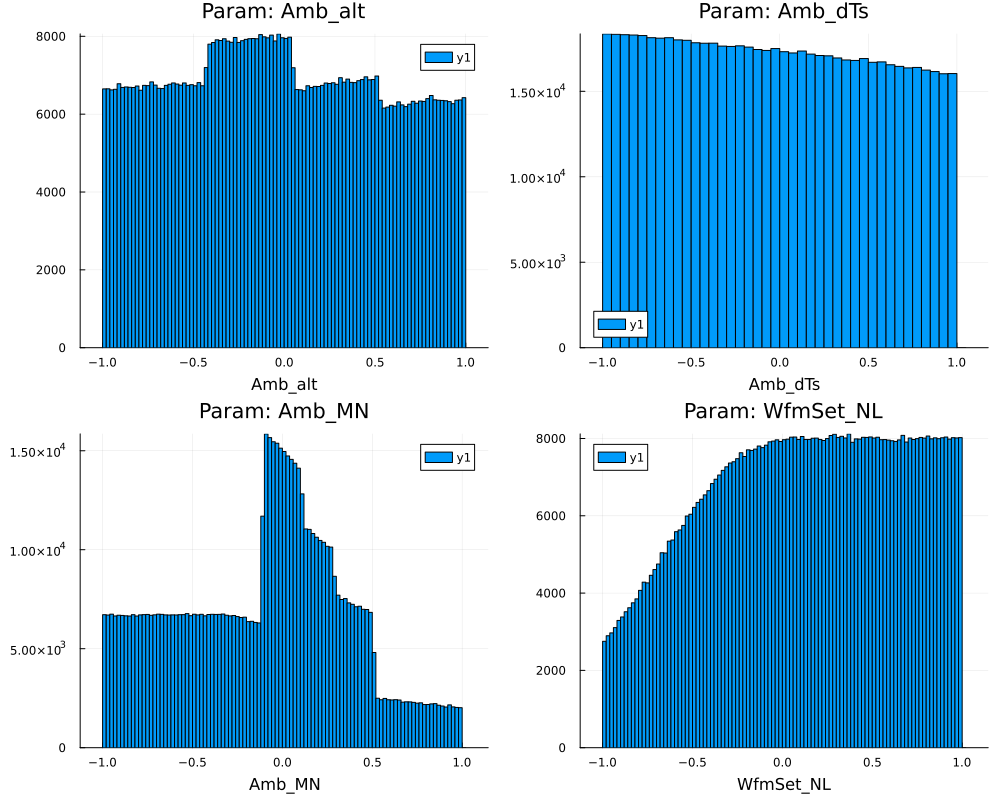}
    \caption{Distribution of the input parameter space (flight envelope)}
    \label{fig:dense_input}
\end{figure}

Active learning of surrogate models has been explored in the aerospace community to retrain surrogate models according to a notion of global error \cite{anantharaman2023approximation, tripathi2024active, zhang2024active}. In this work, active learning is leveraged to balance and normalize the output distribution of the training dataset, by adaptively downsampling a densely sampled input space. This procedure is described in Algorithms \ref{alg:al1} and \ref{alg:al2}. This downsampling is essential to take advantage of deep learning training schemes which rely on assumptions of the distributions of training data in order to be well behaved \cite{naseer2024unbiasednets, bansal2021metabalance}.

\begin{figure}
     \centering
     \begin{subfigure}[b]{0.6\textwidth}
         \centering
         \includegraphics[width=\textwidth]{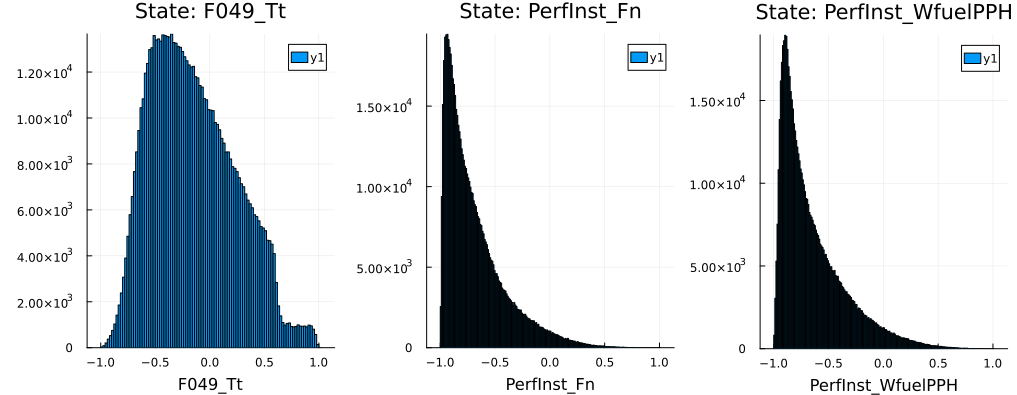}
         \caption{}
         \label{fig:pre_post_al_suba}
     \end{subfigure}
     \hfill
     \begin{subfigure}[b]{0.7\textwidth}
         \centering
         \includegraphics[width=\textwidth]{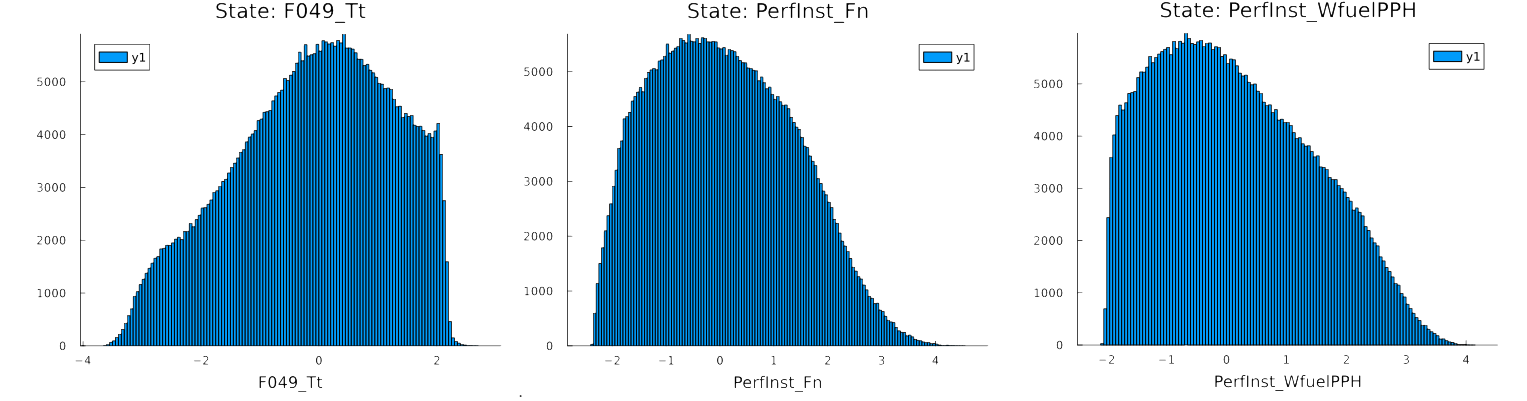}
         \caption{}
         \label{fig:pre_post_al_subb}
     \end{subfigure}
    \caption{Output distribution pre and post adaptive downsampling approach. In \ref{fig:pre_post_al_suba}, the output distribution exhibits skew, whereas the downsampling scheme results in a more uniform output distribution in \ref{fig:pre_post_al_subb}}
    \label{fig:pre_post_al}
\end{figure}

First, a densely sampled set of input space was generated, resulting in 1 million points in the chosen flight envelope. Second, the outputs were computed at each of the million points, resulting in a dense training set. A naive approach would be to use this dense training set for training, whose results we have studied in the following section. Instead, this training set was used to compute the extrema of the outputs. The output space was then sampled using a quasi monte carlo scheme, resulting in 1 million output candidate points. For each of these output points, a nearest neighbors routine was used to compute a best match in the training set, using the Manhattan distance metric \cite{malkauthekar2013analysis}. These best matches were all collected to form a new superior training set. This new training set is, by design, imbued with a more uniformly distributed output space.

The effect of the active learning algorithm is captured in Figure \ref{fig:pre_post_al}. The input parameter space (which is the flight envelope) is densely sampled (over 1 million samples) using latin hypercube sampling \cite{loh1996latin}, producing an input distribution captured in Figure \ref{fig:dense_input}.  The resulting output distribution is not uniform, as shown in Figure \ref{fig:pre_post_al_suba}. In the following section, we shall review that this dataset results in suboptimal performance of the surrogate model. After the adaptive downsampling pass, the output distributions are more uniform across its upper and lower limits, resulting in an output distribution given by \ref{fig:pre_post_al_subb}.

\section{Results and Error Statistics}

The flight envelope parameters are described by Tables \ref{tab:fe1} and \ref{tab:fe2}.

The outputs are as follows: 
\begin{enumerate}
    \item \text{ShH.N} represents the mechanical speed of the shaft 
    \item \text{F049.Tt} represents the turbine temperature.
    \item \text{PerfInst.Fn} represents the net thrust, which is the gross thrust generated by the engine minus the sum of all the drag terms
    \item \text{PerfInst.Wfuel\_PPH} represents the sum of all fuel flows including audit factor.
\end{enumerate}

The specific fuel consumption (SFC) can be calculated as a function of the other outputs, and it is given by a ratio of the net fuel flow and the net thrust. That is,  
$$
\texttt{PerfInst.SFC} = \frac{\texttt{PerfInst.Wfuel\_PPH}}{\texttt{PerfInst.Fn}}
$$
Thus the surrogate model was not trained to predict $\texttt{PerfInst.SFC}$ as an output quantity, so as to calculate it post-facto.

The surrogate model was first trained using the densely sampled input dataset of size 1 million. It was then tested in 1 million test points, resulting in an error distribution given by Table \ref{tab:res_preal}. The results are split into two buckets: what percentage of the test dataset had a prediction with relative error 0.1\% or less, and what percentage had an error of 1\% or less. In particular, the output quantities that had the worst skew (as shown in Figure \ref{fig:pre_post_al_suba}) performed the worst. 

\begin{table}
\caption{\label{tab:table1} Surrogate Model Accuracy pre Active Learning. The percentages indicate what portion of the test dataset fell within a certain accuracy.}
\centering
\begin{tabular}{lcccc}
\hline
Quantity & 0-0.1\% RE & 0.0-1.0\% RE \\\hline
ShH\_N & 98.85\%   & 99.88\% \\
F049\_Tt & 99.78\% & 99.99\% \\
PerfInst\_Fn & 82.60\% & 98.2\% \\
PerfInst\_WfuelPPH & 91.21\% & 99.94\% \\
\hline
\end{tabular}
\label{tab:res_preal}
\end{table}

The surrogate model was then trained using the new downsampled dataset, of size 424,000 points, and tested on 100k test points. The results are captured in Table \ref{tab:res_postal}. Almost all of the test predictions are within 0.1\% relative error, which indicates that the adaptive downsampling approach was effective. 

\begin{table}
\caption{\label{tab:table1} Surrogate Model Accuracy post Active Learning.The percentages indicate what portion of the test dataset fell within a certain accuracy.}
\centering
\begin{tabular}{lcccc}
\hline
Quantity & 0-0.1\% RE & 0.0-1.0\% RE \\\hline
ShH\_N & 99.49\% & 99.90\% \\
F049\_Tt & 100\% & 100\% \\
PerfInst\_Fn & 98.44\% & 99.92\% \\
PerfInst\_WfuelPPH & 99.98\% & 100\% \\
\hline
\end{tabular}
 \label{tab:res_postal}
\end{table}

\begin{figure}
     \centering
     \begin{subfigure}[b]{0.3\textwidth}
         \centering
         \includegraphics[width=\textwidth]{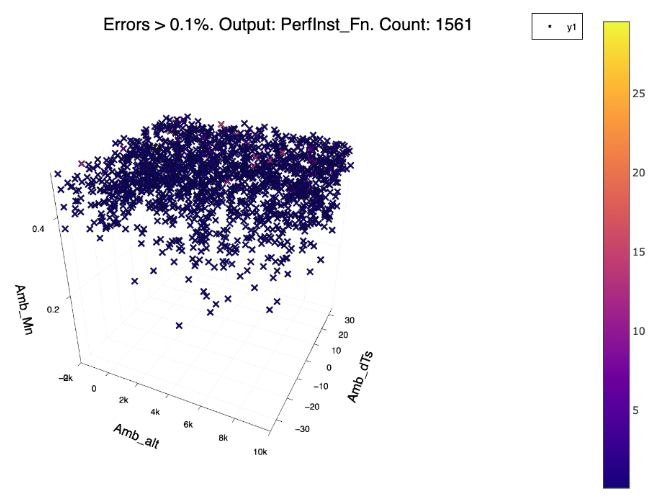}
         \caption{}
         \label{fig:suba}
     \end{subfigure}
     \hfill
     \begin{subfigure}[b]{0.3\textwidth}
         \centering
         \includegraphics[width=\textwidth]{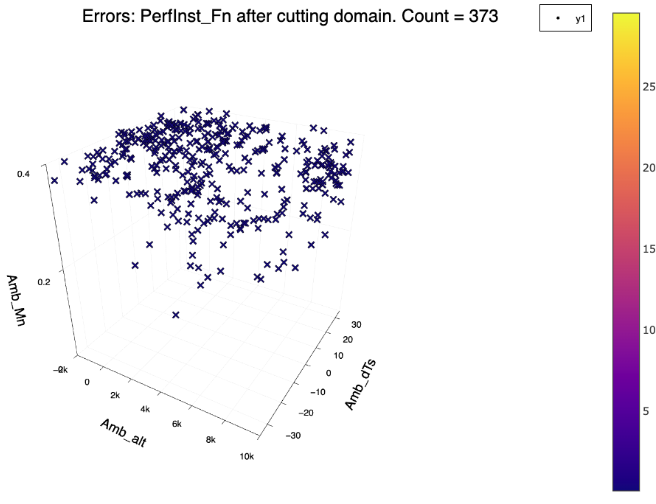}
         \caption{}
         \label{fig:subb}
     \end{subfigure}
     \hfill
     \begin{subfigure}[b]{0.3\textwidth}
         \centering
         \includegraphics[width=\textwidth]{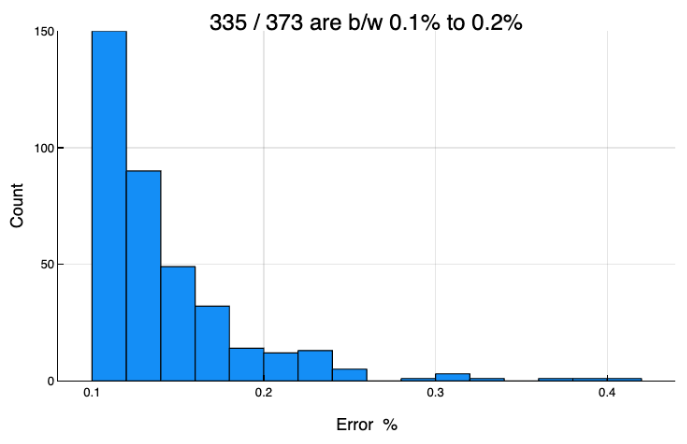}
         \caption{}
         \label{fig:subc}
     \end{subfigure}
        \caption{Distribution of error across input space, considering only points with >0.1\% relative error for the output quantity \texttt{PerfInst\_Fn}. A 3D plot of the error across the flight envelope is shown in \ref{fig:suba}, which indicates that most of the error is concentrated at the boundaries. After cutting the input space by 20\%, i.e., by reducing the mach number input space from (0.0, 0.5) to (0.0, 0.4), the number of high error points reduces by an order of magnitude, as shown in \ref{fig:subb}. Finally, an analyses of the remaining points shows that almost all these remaining points have error between 0.1\% and 0.2\% relative error.}
        \label{fig:error_stats}
\end{figure}

Finally, we conduct an error analysis study on our active learning-based surrogate model, shown in Figure \ref{fig:error_stats}, which is a 3D plot of the error distribution across the flight envelope, considering only points in the test dataset for which relative error is >0.1\%. In effect, Figure \ref{fig:error_stats} conveys that the points of maximum error are largely found in the boundary of the flight envelope, particularly at the boundary of the mach number. The relationship between mach number and altitute, as shown in Table \ref{tab:fe2} is thought to induce a skew during the data-generation process, leading to this behaviour. Thus the surrogate model is seen to be the most sensitive to the mach number. One approach to deal with this is to cut away the edge of the flight envelope at the mach number, as shown in Figure \ref{fig:subb}. This reduces the number of high error points by an order of magnitude, validating the notion that the error is along the boundaries. The final histogram in \ref{fig:subc} shows that of the 300 or so points with relative error > 0.1\%, almost all of them have relative error between 0.1-0.2\%.

\section{Conclusion}
The Digital Echo from JuliaSim, coupled with an active learning algorithm, was used to generate an extremely high accuracy surrogate model with a global relative error of near 0.1\%. We see that for all of the output quantities, greater than 98.4\% of the test set falls within the desired 0.1\% relative error (RE) accuracy metric, and for some quantities, more than $99.7\%$ of the test dataset falls under $0.1\%$ relative error, indicating near-perfect global accuracy.


Most of the high relative error was found at the edge of the mach number. This could be due to the skew in the input distribution when sampling the flight envelope. To combat this skew, another adaptive downsampling pass could be conducted. 

Another approach to generate surrogate models for jet enginees to use an ensemble of different surrogate models \cite{goel2007ensemble} for takeoff, flight and descent, stitched together by a mixture of experts approach \cite{masoudnia2014mixture}. However, for this approach to be feasible, precise segementation of the input space of the physics-based model is required. The authors will leave this approach to future work.

\bibliography{sample}

\begin{thebibliography}{32}
\newcommand{\enquote}[1]{``#1''}
\providecommand{\natexlab}[1]{#1}
\providecommand{\url}[1]{\texttt{#1}}
\providecommand{\urlprefix}{URL }
\expandafter\ifx\csname urlstyle\endcsname\relax
  \providecommand{\doi}[1]{\discretionary{}{}{}https://doi.org/#1}\else
  \providecommand{\doi}[1]{\discretionary{}{}{}\urlstyle{rm}\url{https://doi.org/#1}}\fi

\bibitem[{Yondo et~al.(2018)Yondo, Andr{\'e}s, and Valero}]{yondo2018review}
Yondo, R., Andr{\'e}s, E., and Valero, E., \enquote{A review on design of experiments and surrogate models in aircraft real-time and many-query aerodynamic analyses,} \emph{Progress in aerospace sciences}, Vol.~96, 2018, pp. 23--61.

\bibitem[{Tao and Sun(2019)}]{tao2019application}
Tao, J., and Sun, G., \enquote{Application of deep learning based multi-fidelity surrogate model to robust aerodynamic design optimization,} \emph{Aerospace Science and Technology}, Vol.~92, 2019, pp. 722--737.

\bibitem[{Schaefer et~al.(2020)Schaefer, Romero, Schafer, Leyde, and Denham}]{schaefer2020approaches}
Schaefer, J.~A., Romero, V.~J., Schafer, S.~R., Leyde, B., and Denham, C.~L., \enquote{Approaches for quantifying uncertainties in computational modeling for aerospace applications,} \emph{AIAA Scitech 2020 Forum}, 2020, p. 1520.

\bibitem[{Beran et~al.(2017)Beran, Stanford, and Schrock}]{beran2017uncertainty}
Beran, P., Stanford, B., and Schrock, C., \enquote{Uncertainty quantification in aeroelasticity,} \emph{Annual review of fluid mechanics}, Vol.~49, No.~1, 2017, pp. 361--386.

\bibitem[{Jansson et~al.(2003)Jansson, Nilsson, and Redhe}]{jansson2003using}
Jansson, T., Nilsson, L., and Redhe, M., \enquote{Using surrogate models and response surfaces in structural optimization--with application to crashworthiness design and sheet metal forming,} \emph{Structural and Multidisciplinary Optimization}, Vol.~25, 2003, pp. 129--140.

\bibitem[{Laurenceau and Sagaut(2008)}]{laurenceau2008building}
Laurenceau, J., and Sagaut, P., \enquote{Building efficient response surfaces of aerodynamic functions with kriging and cokriging,} \emph{AIAA journal}, Vol.~46, No.~2, 2008, pp. 498--507.

\bibitem[{Jakobsson and Amoignon(2007)}]{jakobsson2007mesh}
Jakobsson, S., and Amoignon, O., \enquote{Mesh deformation using radial basis functions for gradient-based aerodynamic shape optimization,} \emph{Computers \& Fluids}, Vol.~36, No.~6, 2007, pp. 1119--1136.

\bibitem[{Li et~al.(2019)Li, Kou, and Zhang}]{li2019deep}
Li, K., Kou, J., and Zhang, W., \enquote{Deep neural network for unsteady aerodynamic and aeroelastic modeling across multiple Mach numbers,} \emph{Nonlinear Dynamics}, Vol.~96, 2019, pp. 2157--2177.

\bibitem[{Keane and Voutchkov(2020)}]{keane2020robust}
Keane, A.~J., and Voutchkov, I.~I., \enquote{Robust design optimization using surrogate models,} \emph{Journal of Computational Design and Engineering}, Vol.~7, No.~1, 2020, pp. 44--55.

\bibitem[{Mack et~al.(2007)Mack, Goel, Shyy, and Haftka}]{mack2007surrogate}
Mack, Y., Goel, T., Shyy, W., and Haftka, R., \enquote{Surrogate model-based optimization framework: a case study in aerospace design,} \emph{Evolutionary computation in dynamic and uncertain environments}, 2007, pp. 323--342.

\bibitem[{Williams and Cremaschi(2019)}]{williams2019surrogate}
Williams, B., and Cremaschi, S., \enquote{Surrogate model selection for design space approximation and surrogatebased optimization,} \emph{Computer aided chemical engineering}, Vol.~47, Elsevier, 2019, pp. 353--358.

\bibitem[{Wortmann et~al.(2015)Wortmann, Costa, Nannicini, and Schroepfer}]{wortmann2015advantages}
Wortmann, T., Costa, A., Nannicini, G., and Schroepfer, T., \enquote{Advantages of surrogate models for architectural design optimization,} \emph{AI EDAM}, Vol.~29, No.~4, 2015, pp. 471--481.

\bibitem[{Han et~al.(2012)Han, Zhang et~al.}]{han2012surrogate}
Han, Z.-H., Zhang, K.-S., et~al., \enquote{Surrogate-based optimization,} \emph{Real-world applications of genetic algorithms}, Vol. 343, 2012, pp. 343--362.

\bibitem[{Alizadeh et~al.(2020)Alizadeh, Allen, and Mistree}]{alizadeh2020managing}
Alizadeh, R., Allen, J.~K., and Mistree, F., \enquote{Managing computational complexity using surrogate models: a critical review,} \emph{Research in Engineering Design}, Vol.~31, No.~3, 2020, pp. 275--298.

\bibitem[{Li and Wang(2019)}]{li2019surrogate}
Li, M., and Wang, Z., \enquote{Surrogate model uncertainty quantification for reliability-based design optimization,} \emph{Reliability Engineering \& System Safety}, Vol. 192, 2019, p. 106432.

\bibitem[{Lytle et~al.(2000)Lytle, Follen, Naiman, and Evans}]{lytle2000numerical}
Lytle, J., Follen, G., Naiman, C., and Evans, A., \enquote{Numerical propulsion system simulation (NPSS) 1999 industry review,} Tech. rep., 2000.

\bibitem[{Rackauckas and Abdelrehim()}]{rackauckasscientific}
Rackauckas, C.~V., and Abdelrehim, A., \enquote{Scientific Machine Learning (SciML) Surrogates for Industry, Part 1: The Guiding Questions,} ????

\bibitem[{Rackauckas et~al.(2022)Rackauckas, Gwozdz, Jain, Ma, Martinuzzi, Rajput, Saba, Shah, Anantharaman, Edelman et~al.}]{rackauckas2022composing}
Rackauckas, C., Gwozdz, M., Jain, A., Ma, Y., Martinuzzi, F., Rajput, U., Saba, E., Shah, V.~B., Anantharaman, R., Edelman, A., et~al., \enquote{Composing modeling and simulation with machine learning in Julia,} \emph{2022 Annual Modeling and Simulation Conference (ANNSIM)}, IEEE, 2022, pp. 1--17.

\bibitem[{Bezanson et~al.(2017)Bezanson, Edelman, Karpinski, and Shah}]{bezanson2017julia}
Bezanson, J., Edelman, A., Karpinski, S., and Shah, V.~B., \enquote{Julia: A fresh approach to numerical computing,} \emph{SIAM review}, Vol.~59, No.~1, 2017, pp. 65--98.

\bibitem[{Anantharaman et~al.(2020)Anantharaman, Ma, Gowda, Laughman, Shah, Edelman, and Rackauckas}]{anantharaman2020accelerating}
Anantharaman, R., Ma, Y., Gowda, S., Laughman, C., Shah, V., Edelman, A., and Rackauckas, C., \enquote{Accelerating simulation of stiff nonlinear systems using continuous-time echo state networks,} \emph{arXiv preprint arXiv:2010.04004}, 2020.

\bibitem[{Anantharaman et~al.(2022)Anantharaman, Abdelrehim, Jain, Pal, Sharp, Edelman, Rackauckas et~al.}]{anantharaman2022stably}
Anantharaman, R., Abdelrehim, A., Jain, A., Pal, A., Sharp, D., Edelman, A., Rackauckas, C., et~al., \enquote{Stably accelerating stiff quantitative systems pharmacology models: Continuous-time echo state networks as implicit machine learning,} \emph{IFAC-PapersOnLine}, Vol.~55, No.~23, 2022, pp. 1--6.

\bibitem[{Anantharaman et~al.(2021)Anantharaman, Abdelrehim, Martinuzzi, Yalburgi, Saba, Fischer, Hertz, de~Vos, Laughman, Ma et~al.}]{anantharaman2021composable}
Anantharaman, R., Abdelrehim, A., Martinuzzi, F., Yalburgi, S., Saba, E., Fischer, K., Hertz, G., de~Vos, P., Laughman, C., Ma, Y., et~al., \enquote{Composable and reusable neural surrogates to predict system response of causal model components,} \emph{AAAI 2022 Workshop on AI for Design and Manufacturing (ADAM)}, 2021.

\bibitem[{Roberts et~al.(2022)Roberts, Lara, Henriquez-Auba, Bossart, Anantharaman, Rackauckas, Hodge, and Callaway}]{roberts2022continuous}
Roberts, C., Lara, J.~D., Henriquez-Auba, R., Bossart, M., Anantharaman, R., Rackauckas, C., Hodge, B.-M., and Callaway, D.~S., \enquote{Continuous-time echo state networks for predicting power system dynamics,} \emph{Electric Power Systems Research}, Vol. 212, 2022, p. 108562.

\bibitem[{Anantharaman(2023)}]{anantharaman2023approximation}
Anantharaman, R., \emph{Approximation of Large Stiff Acausal Models}, Massachusetts Institute of Technology, 2023.

\bibitem[{Tripathi et~al.(2024)Tripathi, Kumar, Desale, and Pant}]{tripathi2024active}
Tripathi, M., Kumar, S., Desale, Y.~B., and Pant, R.~S., \enquote{Active Learning-CFD Integrated Surrogate-Based Framework for Shape Optimization of LTA Systems,} \emph{AIAA AVIATION FORUM AND ASCEND 2024}, 2024, p. 4120.

\bibitem[{Zhang et~al.(2024)Zhang, Huang, Shen, Xu, and Niu}]{zhang2024active}
Zhang, H., Huang, W., Shen, Y., Xu, D.-y., and Niu, Y.-b., \enquote{Active learning for efficient data-driven aerodynamic modeling in spaceplane design,} \emph{Physics of Fluids}, Vol.~36, No.~6, 2024.

\bibitem[{Naseer et~al.(2024)Naseer, Prabakaran, Hasan, and Shafique}]{naseer2024unbiasednets}
Naseer, M., Prabakaran, B.~S., Hasan, O., and Shafique, M., \enquote{UnbiasedNets: a dataset diversification framework for robustness bias alleviation in neural networks,} \emph{Machine Learning}, Vol. 113, No.~5, 2024, pp. 2499--2526.

\bibitem[{Bansal et~al.(2021)Bansal, Goldblum, Cherepanova, Schwarzschild, Bruss, and Goldstein}]{bansal2021metabalance}
Bansal, A., Goldblum, M., Cherepanova, V., Schwarzschild, A., Bruss, C.~B., and Goldstein, T., \enquote{MetaBalance: high-performance neural networks for class-imbalanced data,} \emph{arXiv preprint arXiv:2106.09643}, 2021.

\bibitem[{Malkauthekar(2013)}]{malkauthekar2013analysis}
Malkauthekar, M., \enquote{Analysis of euclidean distance and manhattan distance measure in face recognition,} \emph{Third International Conference on Computational Intelligence and Information Technology (CIIT 2013)}, IET, 2013, pp. 503--507.

\bibitem[{Loh(1996)}]{loh1996latin}
Loh, W.-L., \enquote{On Latin hypercube sampling,} \emph{The annals of statistics}, Vol.~24, No.~5, 1996, pp. 2058--2080.

\bibitem[{Goel et~al.(2007)Goel, Haftka, Shyy, and Queipo}]{goel2007ensemble}
Goel, T., Haftka, R.~T., Shyy, W., and Queipo, N.~V., \enquote{Ensemble of surrogates,} \emph{Structural and Multidisciplinary Optimization}, Vol.~33, 2007, pp. 199--216.

\bibitem[{Masoudnia and Ebrahimpour(2014)}]{masoudnia2014mixture}
Masoudnia, S., and Ebrahimpour, R., \enquote{Mixture of experts: a literature survey,} \emph{Artificial Intelligence Review}, Vol.~42, 2014, pp. 275--293.

\end{thebibliography}

\end{document}